# Tailored joint fabrication process derived ultra-low resistance MgB$_2$ superconducting joint

Dipak Patel[a], Akiyoshi Matsumoto[a,*], Hiroaki Kumakura[a], Gen Nishijima[a], Minoru Maeda[b], Su-Hun Kim[c], Seyong Choi[b], Jung Ho Kim[d]

[a] National Institute for Materials Science (NIMS), 1-2-1 Sengen, Tsukuba, Ibaraki 305-0047, Japan

[b] Department of Electrical Engineering, Kangwon National University, Kangwon 25913, Republic of Korea

[c] Department of Electrical Engineering, Kyungpook National University, Daegu 41566, Republic of Korea

[d] Institute for Superconducting and Electronic Materials, Australian Institute for Innovative Materials, University of Wollongong, North Wollongong, New South Wales 2500, Australia

## Abstract

We report an ultra-low resistance superconducting joint using unreacted multifilament MgB$_2$ wires produced by tailoring the powder compaction pressure within the joint with heat treatment conditions. The joint demonstrated an ultra-low resistance of $5.48 \times 10^{-15}$ Ω and critical current ($I_c$) of 91.3 A at 20 K in self-field. The microstructural and composition studies of the joint revealed cracks and a high amount of MgO, respectively. These two features reduced the $I_c$ of the joint to some extent; nevertheless, the joint resistance was not affected by it. Our tailored joining process will play a pivotal role in superconducting joint development.

*Corresponding author e-mail address: A. Matsumoto (matsumoto.akiyoshi@nims.go.jp)

Since the discovery of magnesium diboride ($MgB_2$) in 2001 as a superconducting material with a critical temperature of 39 K, its performance has been greatly enhanced by means of focused experimental efforts and developing a fundamental understanding of this material [1–10]. In the meantime, the $MgB_2$ material has also been seamlessly transformed into the high-performance wire and tape conductors for practical applications [11–14]. In particular, these $MgB_2$ conductors are very attractive for liquid helium (LHe)-free magnetic resonance imaging (MRI) magnet application, because, in recent years, the price of LHe is skyrocketing and there is a global crisis of helium, which has made maintenance of MRI systems even more expensive in many parts of the world [15–17]. However, to produce an ultra-stable magnetic field of the desired decay rate of <0.1 ppm $h^{-1}$ in an $MgB_2$ magnet for MRI application, *'superconducting joints'* are essential [18]. The superconducting joints allow the magnet to form a superconducting closed-loop by short-circuiting two ends of the magnet, once the magnet is charged with the desired current. Thereby, current flows in the closed-loop without any noticeable decay and keeps producing an ultra-stable magnetic field. This operation is called the persistent-mode operation [19].

Superconducting joints using monofilament $MgB_2$ conductors have been widely reported [12,20–28]. It is essential, however, to use multifilament $MgB_2$ conductors in the MRI magnet to reduce the AC losses [29,30]. The first joint using multifilament $MgB_2$ tape was reported by Li et al. [25]. Their heated-twice joint attained critical current density ($J_c$) of $1.74 \times 10^5$ A $cm^{-2}$ at 4.2 K in 1 T and joint resistance of $6.24 \times 10^{-11}$ $\Omega$ at 4.2 K in self-field [31]. Yao et al. fabricated several joints using multifilament $MgB_2$ wires [21]. One of their joints achieved critical current ($I_c$) of 200 A at 10 K in self-field. Nardelli et al. demonstrated a multifilament superconducting joint in a closed-loop coil carrying 300 A transport current and joint resistance of <$10^{-14}$ $\Omega$ at 20 K in self-field [32]. Park et al. reported a maximum $I_c$ of 103 A at 10 K using multifilament $MgB_2$ wire [22]. Very recently, Yoo et al. fabricated several superconducting joints using multifilament $MgB_2$ wire by varying the powder compaction pressure and angle of the filaments cutting [33]. Their best joint achieved $I_c$ of 262 A at 20 K in self-field and joint resistance of <$1.5 \times 10^{-14}$ $\Omega$ at 20 K in self-field. Despite the fact that superconducting joints will be some of the vital components in the $MgB_2$ MRI magnet system, the research on the superconducting joining technology using multifilament $MgB_2$ conductors is still in the infancy stage. Hence, there is a further need to improve the joining technology and associated processes. In particular, if the superconducting joints using multifilament $MgB_2$ conductor can be fabricated with further lower resistance, it will be certainly beneficial to improve magnetic field stability during the persistent-mode operation in the $MgB_2$ MRI magnet [34]. Moreover, surprisingly, no superconducting joint using a multifilament $MgB_2$ conductor has ever been evaluated in a magnetic field above 4.2 K temperature. In a practical scenario, however, the joints will be placed in the MRI magnet system under some magnetic field. Herein, therefore, we report superconducting joints using unreacted multifilament $MgB_2$ wires and their evaluation results in different magnetic fields and temperatures. The powder compaction pressure within the joint and heat treatment conditions were tailored to realize high performing superconducting interfaces between the filaments of two $MgB_2$ wires.

The unreacted multifilament $MgB_2$ wire was fabricated using an *in situ* process and supplied by Sam Dong Co., Ltd [35,36]. Figure S1 in supplementary material shows the optical cross-sectional image of the $MgB_2$ wire and Table S1 shows its specifications. The transport $I_c$(s) of the joints were measured in a variable temperature insert (VTI) of a superconducting magnet with dc current up to 200 A at different temperatures and magnetic fields using the four-probe method with the criterion of 1 $\mu V\ cm^{-1}$. A closed-loop coil was fabricated using one superconducting joint to measure the joint resistance using the field-decay method [20]. The photograph of the $MgB_2$ closed-loop coil and its test setup are shown in Figure S2, whereas specifications of the coil are shown in Table S2. The cross-section of the joint was observed in a Hitachi SU-70 scanning electron microscope (SEM). For analyzing the powder specimen obtained from the $MgB_2$ bulk of the joint, the X-ray diffraction (XRD) pattern was acquired using Rigaku Miniflex II. The quantitative analysis was performed using a RIETAN-FP Rietveld refinement package [37].

Figure 1(a) shows our superconducting joint fabrication process for unreacted multifilament $MgB_2$ wires. In preparation, Monel and copper (Cu) from the end of the unreacted wires were removed using nitric acid (60% w/w, 1 h). The joint fabrication process involved, wire cutting at 20° angle, aligning wires in stainless steel (SS) mould, filling magnesium (Mg) and boron (B) powders in stoichiometric ratio for $MgB_2$ in the mould cavity, compacting or pressing the powder mixture, applying the high-temperature sealing material (Durabond[TM] 7032), and heat treatment of the joint in an inert atmosphere. The temperature ramp rate during the heat treatment was 5 °C min$^{-1}$ and the cooling down rate was 10 °C min$^{-1}$. The schematic illustration of the joint when Mg + B power filled in the mould cavity and two-way solid-state diffusion of Mg during heat treatment at 650 °C between Mg + B powder of filaments and cavity is shown in Figure 1(b). As shown via illustration in Figure 1(c), after the heat treatment, $MgB_2$ was formed in the filaments and the cavity along with several voids due to the diffusion of Mg into B sites. The SEM image of one the filaments after heat treatment inside the joint part is shown in Figure 1(d). The superconducting interface was formed between $MgB_2$ filament core and $MgB_2$ bulk as shown in the figure.

Figure 2(a) shows compaction pressure dependence on the $I_c$ at 20 K in the self-field of the joints heat-treated at two different temperatures. As can be seen in the figure, the joint heat-treated at 650 °C for 30 min and compaction pressure of 20 kN showed the highest $I_c$ of 91.3 A among the joints with the same heat treatment condition. On the other hand, the joint heat-treated at 670 °C for 60 min and compaction pressure of 30 kN showed the highest $I_c$ of 97.7 A among the joints with the same heat treatment condition. These results suggest that to achieve superior transport performance of the joints, the compaction pressure needs to be tailored with heat treatment conditions. Based on the in-field performance evaluation of the two high performing joints, the joint heat-treated at 650 °C for 30 min was further studied in detail in this work. Magnetic field dependence of the critical current at different temperatures of the selected joint is shown in Figure 2(b). To the best of our knowledge, this work is first to evaluate a superconducting joint of multifilament $MgB_2$ conductors in a magnetic field above 4.2 K temperature. As can be seen in the figure, the joint achieved maximum $I_c$ of 193.5 A at 10 K in self-field. On the other hand, in 1 T field, $I_c$ values of the joint were 43.7 A, 63.5 A, and 85.5 A, at 20 K, 15 K, and 10 K, respectively. The voltage versus current characteristics of the joint at different temperatures in 1 T is shown in Figure S3. The $I_c$ performance of our joint at 10 K in self-field is comparable to the joint results reported by Yao et al [21]. However, it is lower at 20 K in self-field compared with $I_c$ of the joint recently reported by Yoo et al. [33]. Therefore, we further need to improve the transport $I_c$ of our joint.

Following the evaluation of the transport property of the joint, field-decay measurement of the joint was conducted to precisely measure the joint resistance. As shown in Figure 3(a), for trapping the magnetic field in the closed-loop coil, firstly, the background field was increased to 10 kG when the joint temperature in the coil reached about 55 K during cool down. Once the joint temperature in the coil was stabilized at about 20 K, the background field was slowly reduced to zero. During this process, some magnetic field was trapped in the closed-loop as shown in Figure 3(b). The trapped field was allowed to stabilized up to 0.66 h of the experiment time. At this time, the trapped field at the hall sensor location was 13.69 G, which was equivalent to approximately 22.4 A trapped current according to finite element analysis calculation. The trapped field of the closed-loop coil was observed for a little over 13.5 h (Figure 3(c)). At the time of starting persistent-mode operation, the closed-loop coil field was 13.69 G ($t_1$: 2400 sec), which was decayed to 13.60 G ($t_2$: 51053 sec) in about 13.51 h (t:48653 sec). The inductance of the closed-loop coil was 0.0404 µH. The total calculated joint resistance according to $L - R$ circuit decay equation was $5.48 \times 10^{-15}$ Ω at 20 K in self-field [38]. The magnetic field decay line calculated using the $L - R$ circuit time constant is also shown in Figure 3(c) using the dashed line, which matches very well with the actual decay. The joint resistance reported in this work is the lowest joint resistance reported to date for any $MgB_2$ joint at 20 K in self-field. After observing persistent-mode for enough duration, the temperature of the coil was increased along with the joint. As can be seen in Figure 3(d), when the temperature on the joint was started to increase, the trapped field of the closed-loop coil was started to decrease and eventually reached zero.

To correlate the transport property and ultra-low resistance of our joint with its microstructure, the cross-section of the joint was analyzed using SEM. Figure 4(a) shows the cutting location at which the joint was cut to observe the cross-section area of the joint. The SEM cross-sectional image of the joint is shown in Figure 4(b). To further observe the interface between one of the filaments and $MgB_2$ bulk, a higher magnification SEM image was taken from the selected location shown in Figure 4(b). As can be seen in Figure 4(c), the core of the $MgB_2$ filament was very dense; however, the bulk of the $MgB_2$ had so many pores in it. The pore size in the $MgB_2$ bulk is generally dominated by the size of the initial Mg powder. This is because, during the solid-state reaction, Mg particles diffuse into B particles and form $MgB_2$ and leaving pores behind. Generally, it is difficult to avoid pores completely from the $MgB_2$ bulk obtained using Mg and B powders. As shown in Figure 4(c), we also observed some cracks in the $MgB_2$ bulk as shown using yellow arrows. The cracks might have generated due to over compaction pressure in the joint. Further tailoring of compaction pressure around 20 kN force might reduce the cracks in the joint part. These cracks can be attributed to somewhat lowering $I_c$ of our joint compared with Yoo et al. [33]. The interface between the $MgB_2$ filament and bulk appeared very good with some half pores on the edge. As illustrated in Figure 1(b), two-way solid-state diffusion of the Mg particles between the neighboring Mg + B powder and the core of the filaments inside the joint would have taken place during heat treatment at 650 °C to form a superconducting interface between them. This can be attributed to achieving true superconducting joint and ultra-low resistance in our joint.

We used the high-temperature sealing material at the top and side of our joint to reduce the evaporation of Mg during the heat-treatment process. This is because the melting point of Mg is about 650 °C. If we did not use the sealing material, the evaporation of Mg would have led to a stoichiometric imbalance in our $MgB_2$ bulk, thereby, further reducing the transport performance of the joint. However, the use of the sealing material also brought the risk of the reaction of the sealing material with the bulk of the joint. To clarify this point further, we carried out an XRD analysis of the powder specimen obtained from the joint's MgB$_2$ bulk. Figure 4(d) shows the XRD pattern of the powder specimen. As can be seen in the figure, all the major Bragg reflections either identified as MgB$_2$ or magnesium oxide (MgO). No other major reflections were observed, which implies that the sealing material did not react with the MgB$_2$ bulk. Nevertheless, the MgO content in the MgB$_2$ bulk was very high at about 26.3 wt.%. This suggests that the Mg of the premix powder was either contaminated or Oxygen might be introduced in the bulk during fabrication or heat-treatment process. In our previous work, we observed an MgO fraction of about 13.3 wt.% in the MgB$_2$ bulk of the monofilament MgB$_2$ joint [20]. The secondary phase MgO generally exists at the MgB$_2$ grain boundary, thereby affects the grain connectivity [39]. A higher fraction of MgO in the joint's MgB$_2$ bulk, therefore, might be the other reason for lowering $I_c$ of our joint. Interestingly, a higher fraction of MgO does not seems to affect the joint resistance, however.

In summary, we have presented an ultra-low resistance superconducting joint using multifilament MgB$_2$ wires by tailoring the joint fabrication process and its evaluation results in detail. Our rationally fabricated joint using 650 °C for 30 min heat treatment and 20 kN compaction pressure attained $I_c$ of 193.5 A at 10 K in self-field and 43.7 A at 20 K in 1 T. This work is first to evaluate any superconducting joints using multifilament MgB$_2$ conductors in a magnetic field above 4.2 K temperature. The precisely measured joint resistance using the field-decay method was 5.48 × 10$^{-15}$ Ω at 20 K in self-field. This value is the lowest joint resistance reported value to date for any MgB$_2$ joint. The SEM observation of the joint cross-section showed some cracks and the XRD analysis of the powder specimen obtained from the MgB$_2$ bulk revealed a very high MgO impurity content of 26.3 wt.%. These two features were attributed to lowering $I_c$ of the joint. Surprisingly, cracks in the joint area or excessive MgO impurity did not affect the joint resistance. Our findings provide insight for interface engineering between MgB$_2$ superconducting filaments to produce superconducting joints with superior properties for practical applications.

## Acknowledgements

This work was supported by the Japan Society of the Promotion of Science (JSPS) KAKENHI Grant Number JP18F18714, and Cryogenic Station, Research Network and Facility Services Division, National Institute for Materials Science (NIMS), Japan. Dipak Patel is a JSPS

International Research Fellow. This work was also partially supported by the Basic Science Research Program through the National Research Foundation of Korea funded by the Ministry of Education (NRF-2017R1D1A3B03035092) and Technology Innovation Program or Industrial Strategic Technology Development Program (20002088, Scalable integration of $MgB_2$ superconducting wire towards cost-effectiveness and industrial competitiveness) funded by the Ministry of Trade, Industry & Energy (MOTIE, Korea).

**Figure captions**

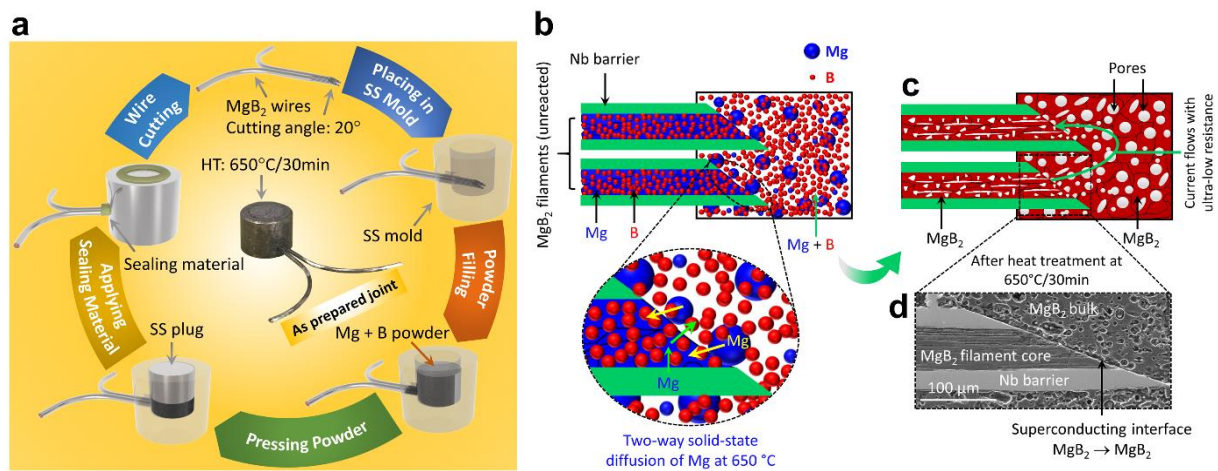

**Figure 1.** (a) Superconducting joint fabrication procedure for an unreacted multifilament $MgB_2$ wires. The photograph of the completely fabricated joint is shown in the center. HT means a heat treatment condition. (b) schematic illustration of the joint when Mg + B power filled in the mould cavity and two-way solid-state diffusion of Mg during heat treatment at 650 °C

between Mg + B powder of filaments and cavity. (c) illustration of the joint after the heat treatment. (d) SEM image of one the filament after heat treatment inside the joint.

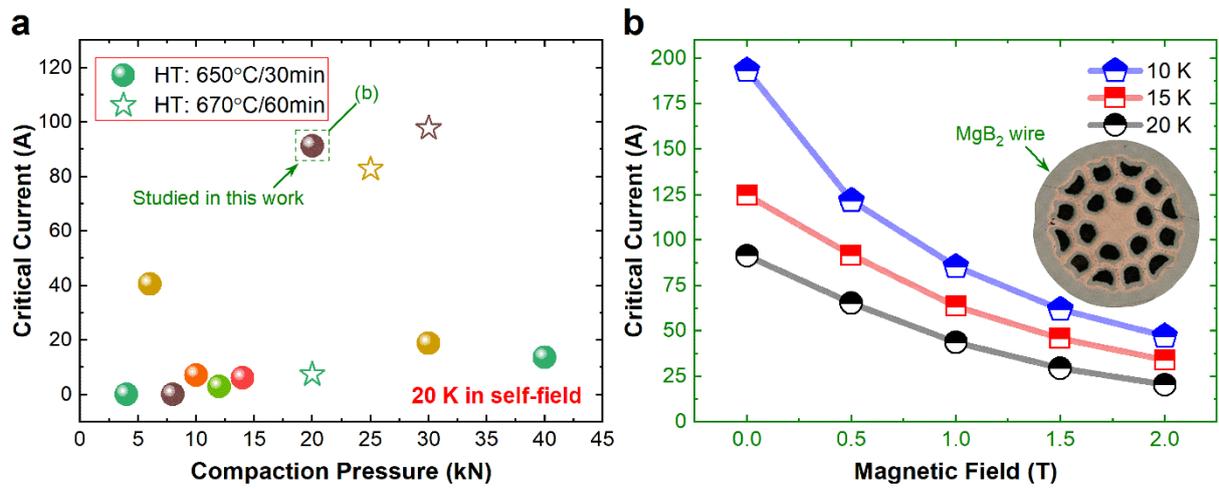

**Figure 2.** (a) Critical current versus compaction pressure characteristics of the joints heat-treated at 650 °C for 30 min and 670 °C for 60 min. (b) Magnetic field dependence of the critical current at different temperatures of the selected joint. The cutting angle of the filaments in all the joints was 20°. HT means a heat treatment condition.

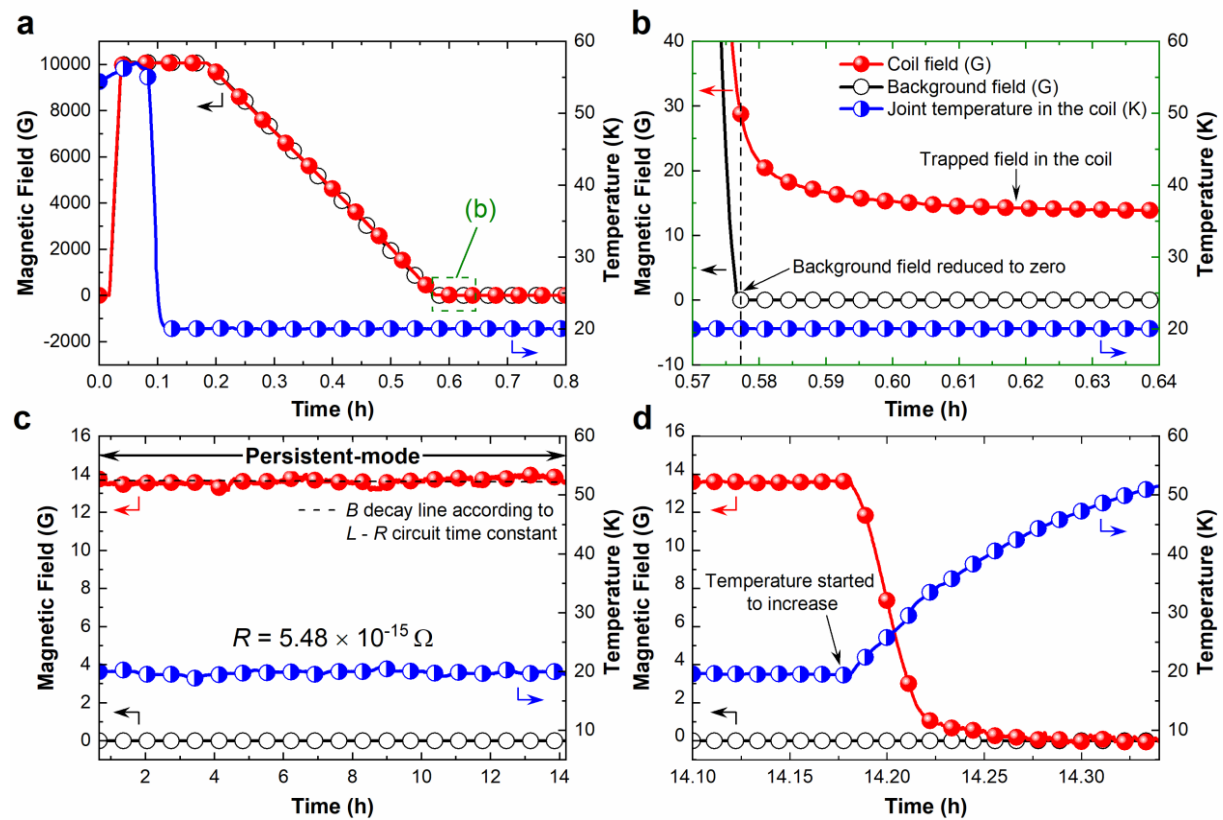

**Figure 3.** Plots of the field-decay measurement of the MgB$_2$ closed-loop coil. All the plots show the magnetic field and temperature versus time characteristics of the closed-loop coil. (a) During the background field reducing to zero. (b) A trapped field in the closed-loop coil when the background field reduced to zero. (c) Persistent-mode data of the closed-loop coil including $B$ decay line calculated according to $L - R$ circuit time constant. (d) The temperature of the closed-loop coil was increased and the trapped field started to decrease. The coil means the closed-loop coil. The scale in figure c is from 0.66 h to 14.18 h. The legends in all the figures are the same.

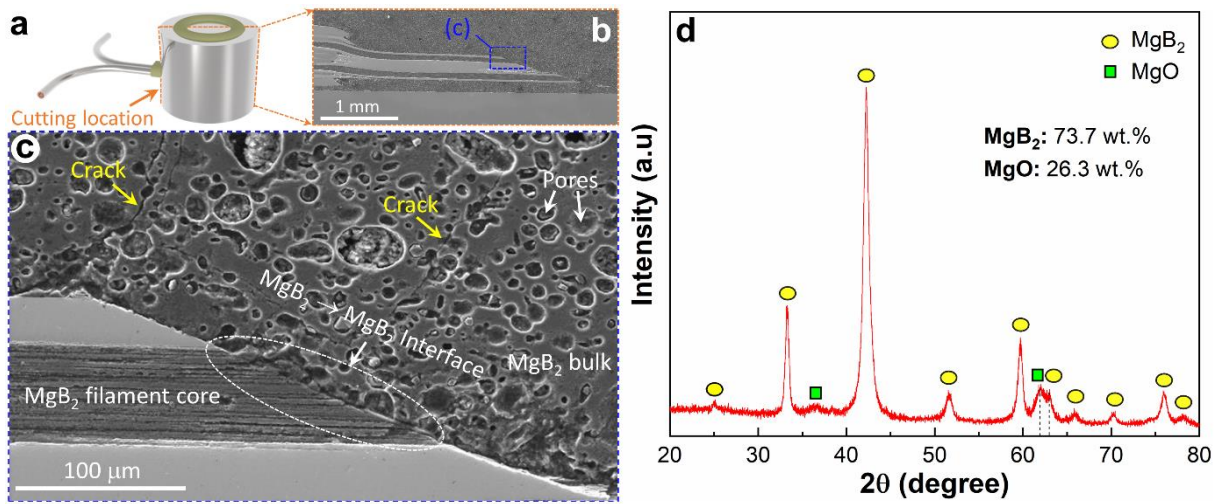

**Figure 4.** (a) The cutting location of the joint. (b) SEM cross-sectional image of the joint. (c) Magnified SEM cross-sectional image of the selected location of figure (b). The interface between one of the filaments of the MgB$_2$ filament core and bulk MgB$_2$ is shown by an elongated circle. Cracks in the MgB$_2$ bulk are shown by yellow arrows and descriptions. (d) XRD pattern of the powder specimen obtained from the MgB$_2$ bulk of the joint.

# Supplementary Material

**Tailored joint fabrication process derived ultra-low resistance MgB$_2$ superconducting joint**

Dipak Patel[a], Akiyoshi Matsumoto[a,*], Hiroaki Kumakura[a], Gen Nishijima[a], Minoru Maeda[b], Su-Hun Kim[c], Seyong Choi[b], Jung Ho Kim[d]

[a] National Institute for Materials Science (NIMS), 1-2-1 Sengen, Tsukuba, Ibaraki 305-0047, Japan

[b] Department of Electrical Engineering, Kangwon National University, Kangwon 25913, Republic of Korea

[c] Department of Electrical Engineering, Kyungpook National University, Daegu 41566, Republic of Korea

[d] Institute for Superconducting and Electronic Materials, Australian Institute for Innovative Materials, University of Wollongong, North Wollongong, New South Wales 2500, Australia

*Corresponding author e-mail address: A. Matsumoto (matsumoto.akiyoshi@nims.go.jp)

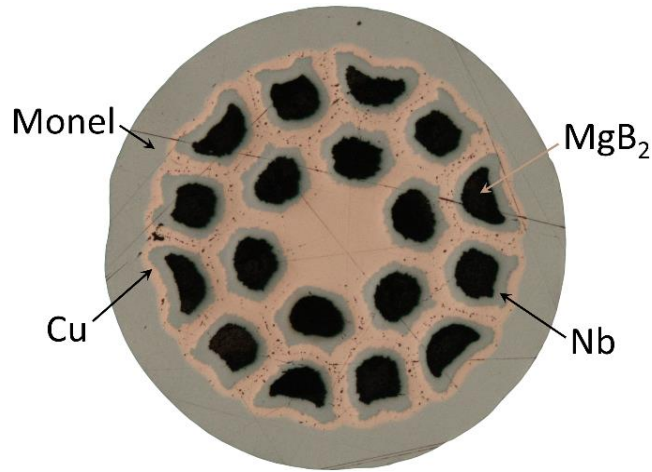

**Figure S1.** Optical cross-sectional image of the MgB$_2$ wire. The diameter of the wire is 0.96 mm.

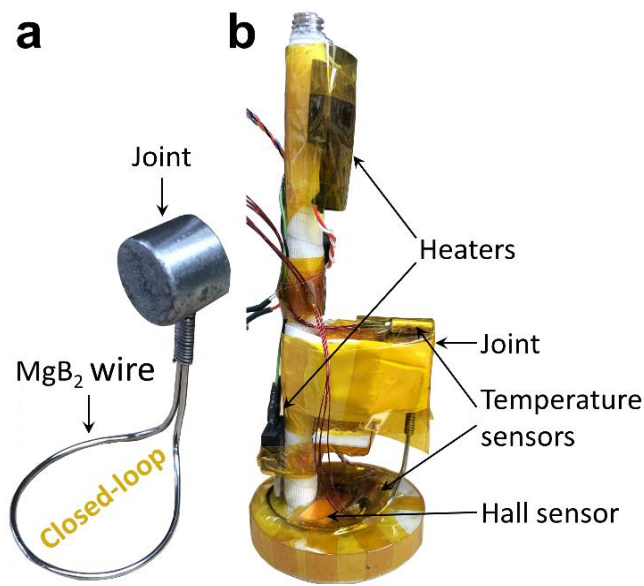

**Figure S2.** (a) Photograph of the MgB$_2$ closed-loop coil. (b) Photograph of the MgB$_2$ closed-loop coil installation on the test probe for the field-decay measurement. Hall sensor was installed at the known location inside the closed-loop coil. One temperature sensor was installed near the MgB$_2$ closed-loop and one on the joint. Two heaters (top and bottom) were installed to control the temperature of the closed-loop coil.

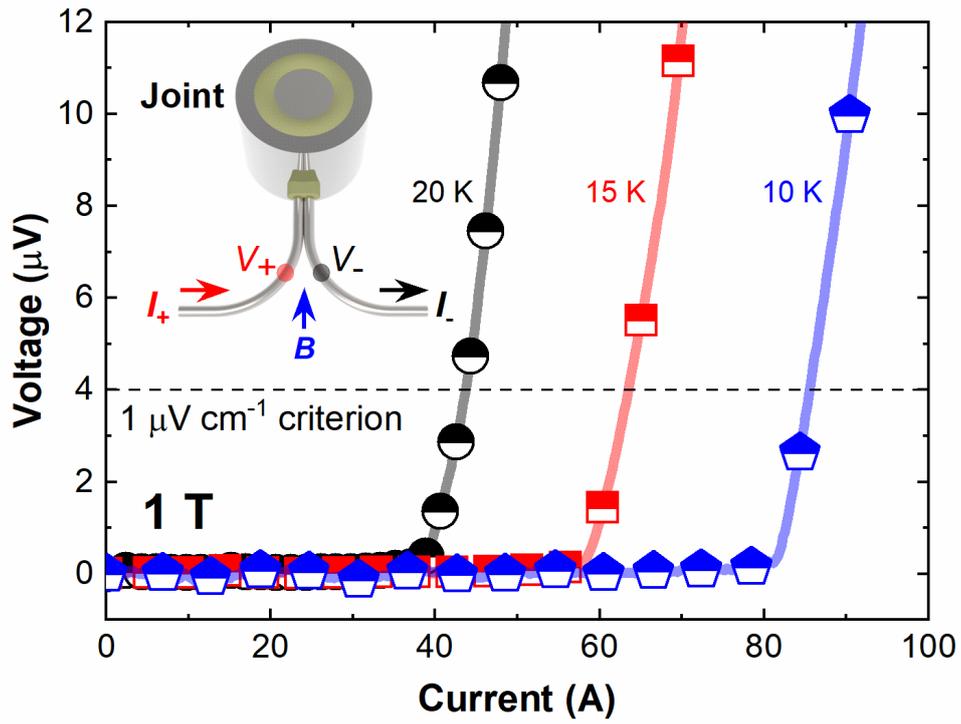

**Figure S3.** Voltage versus current characteristics of the joint at different temperatures in 1 T. The distance between voltage taps on the joint was 4 cm.

**Table S1.** Specifications of the MgB$_2$ superconducting wire. The $I_c$ data was measured when the wire was heat-treated at 650 °C for 30 min.

| Specifications | Values |
|---|---|
| Superconductor | MgB$_2$ |
| Doping | No |
| Processing type | *In situ* |
| Sheath, barrier, stabilizer | Monel (Cu + Ni), Nb, Cu |
| Diameter | 0.96 mm |
| Number of filaments | 18 + 1 (Cu) |
| MgB$_2$ filling factor | 15% |
| Cu fraction | 25% |
| $I_c$ (A) 2 T, 20 K | 86 A |
| Manufacturer | Sam Dong, Korea |

**Table S2.** Specifications of the closed-loop coil.

| Specifications | Values |
|---|---|
| Coil type | Close-loop |
| Superconductor | MgB$_2$ (18 + 1) |
| Wire diameter | 0.96 mm |
| Coil inner/outer diameter | 19 mm, 20.92 mm |
| Calculated parameters using FEA | |
| Inductance | 0.0404 µH |
| Trapped field (at hall sensor) | 13.69 G |
| Trapped current | 22.4 A |